\documentstyle[epsf,a4]{article}  
\begin{document}                
\newcommand{\manual}{rm}        
\newcommand\bs{\char '134 }     
\newcommand{\Het}{$^3{\mathrm{He}}$}
\newcommand{\Hef}{$^4{\mathrm{He}}$}
\newcommand{\A}{{\mathrm{A}}}
\newcommand{\D}{{\mathrm{D}}}
\newcommand{\simlt}{\stackrel{<}{{}_\sim}}
\newcommand{\simgt}{\stackrel{>}{{}_\sim}}
\newcommand{\keV}{\;{\mathrm{keV}}}
\newcommand{\MeV}{\;{\mathrm{MeV}}}
\newcommand{\TeV}{\;{\mathrm{TeV}}}
\newcommand{\GeV}{\;{\mathrm{GeV}}}
\newcommand{\eV}{\;{\mathrm{eV}}}
\newcommand{\ergs}{\;{\mathrm{ergs}}}
\newcommand{\cm}{\;{\mathrm{cm}}}
\newcommand{\s}{\;{\mathrm{s}}}
\newcommand{\sr}{\;{\mathrm{sr}}}
\newcommand{\lab}{{\mathrm{lab}}}
\newcommand{\ts}{\textstyle}
\newcommand{\ol}{\overline}
\newcommand{\be}{\begin{equation}}
\newcommand{\ee}{\end{equation}}
\newcommand{\ba}{\begin{eqnarray}}
\newcommand{\ea}{\end{eqnarray}}
\newcommand{\rau}{\rho_{\mathrm Au}}
\newcommand{\nn}{\nonumber}
\newcommand{\N}{{\mathrm{N}}}
\newcommand{\css}{({\mathrm{cm}}^2-{\mathrm{s}}-{\mathrm{sr}})^{-1}}
\newcommand{\pp}{$\overline{p}(p)-p\;\;$}
\renewcommand{\floatpagefraction}{1.}
\renewcommand{\topfraction}{1.}
\renewcommand{\bottomfraction}{1.}
\renewcommand{\textfraction}{0.}               
\renewcommand{\thefootnote}{F\arabic{footnote}}
\title{Very high-energy neutrinos from slowly decaying, massive
dark matter, as a source of explosive energy for gamma-ray bursts}
\author{Saul Barshay and Georg Kreyerhoff\\
III. Physikalisches Institut\\
RWTH Aachen\\D-52056 Aachen\\Germany}
\maketitle
\begin{abstract}                
We consider a speculative model for gamma-ray bursts (GRB), which
predicts that the total kinetic energy in the ejected matter is less
than the total energy in the gamma rays. There is also secondary
energy in X-rays, which are emitted contemporaneously with
the gamma rays. The model suggests that bremsstrahlung and Compton
up-scattering by very energetic electrons, are important processes
for producing the observed burst radiation. The dynamics naturally allows
for the possibility of a moderate degree of beaming of matter
and radiation in some gamma-ray bursts. GRB are predicted to have
an intrinsically wide distribution in total energies, in
particular, on the low side. They are predicted to occur out to 
large red-shifts, $z\sim 8$, in local regions of dense matter.
\end{abstract}
\section{Introduction}
The most striking empirical aspect of cosmological gamma-ray bursts (GRB)
is the explosive release of a large total energy in energetic photons
in a very short time interval. Of the order of $10^{52}$ ergs may
emerge explosively outward from the source in less than one minute,
in the form of gamma rays of $\sim 0.1\MeV$ to $\sim 1\MeV$, and X-rays.
The dynamical origin of such explosive emission remains a mystery, at
present. This is so, despite the fact that a number of hypotheses \cite{ref1,ref2,ref3}
invoke the gravitational energy released by infalling matter in the vicinity
of a black hole, or during the formation of a black hole, as the principal
source of energy. Models of this type predict that a comparably large
amount of energy is in the kinetic energy of matter (baryons), which
is explosively ejected outwards, as the result of some unknown dynamical
mechanism. This essential prediction is not unambiguously verified by the data
on gamma-ray bursts, at present \cite{ref4}. Indeed, there is analysis \cite{ref5} of
data which suggests that the kinetic energy in ejecta may be less than the
energy in gamma rays. The purpose of this paper is to present some general
results of the hypothesis that the primary source of the energy observed in 
gamma rays arises from the energy in extremely high-energy neutrinos (and
antineutrinos) which accumulate over long time intervals, in the immediate
vicinity of a very massive, central body which is composed of massive
particles of dark matter (inflatons) \cite{ref6,ref7}.\footnote{The
inflaton mass has been calculated in Ref.~6 to be near to $10^{11}\GeV$.
The lifetime has been estimated to be $\tau\sim 10^{25}\s$.}
These neutral, scalar quanta may be metastable \cite{ref6}. $^{\mathrm F1}$
They may decay into neutrino-antineutrino pairs, but with a very long
lifetime, which has been estimated to be many orders of magnitude greater
than the present age of the universe, because the decay matrix element
is proportional to a (relatively) tiny neutrino mass \cite{ref6}. It
is now established that neutrinos have mass \cite{ref8}. \footnote{The largest mass,
probably that of $\nu_\tau$, is possibly of the order of 0.06 eV.\cite{ref9}
The other, lighter neutrino masses may not be less than a factor of 0.1 
smaller.\cite{ref8} Flavor mixing occurs \cite{ref8,ref9}; we do not deal
with this explicitly.}
The essential assumption is that the very energetic neutrinos which are present
from decaying inflatons, can acquire momentum components which allow them to
circulate,
and accumulate in a spherical envelope, near to the massive central body, 
which is assumed to be near
to the condition of a black hole \cite{ref10}. \footnote{The mass $M$, 
and the radius $r$, of the central body
are assumed to be approximately related by $r\simgt r_{\mathrm{S}} = 2GM$,
where $G$ denotes the gravitational constant, and $r_{\mathrm{S}}$ is the
Schwarzschild radius. For a black hole, a massless
photon has a metastable orbit at $3r_{\mathrm{S}}/2$; particles with mass
have orbits out from $3r_{\mathrm{S}}$.\cite{ref10} }
The relevant
decaying inflatons may be circulating in the immediate vicinity of the central body.
In an appendix, we give a heuristic argument that such an accumulation might
be possible. For the moment, we wish to go directly to the consequences
of a rapid encounter of these energetic neutrinos with a sufficiently 
dense toroidal configuration of infalling matter. We show that this 
can give rise to a gamma-ray burst,
within a naturally occurring total time interval of less than 2 minutes,
and with irregular fluctuations in time intervals of a natural order of
magnitude of less than 1 second. These time intervals are calculated below, from
the macroscopic and the microscopic dynamics. Contrary to the above-mentioned
prediction, we estimate that the total kinetic energy in the ejecta is likely
to be significantly less than the total energy in the gamma rays. Gravitational
energy released by infalling matter is present, but it is secondary. We estimate
the contribution as a source of X-rays, which are emitted contemporaneously
with the gamma rays. The specific dynamics predicts an intrinsically wide
distribution of observed total energy in gamma rays, in particular, for
bursts at energies below $\sim 10^{52}$ ergs. The dynamics suggests that
gamma-ray bursts have been prolific at very early times in the universe, out
to red-shifts of the order of 8, near to the times of activation of the first
``engines'' in  active galactic nuclei \cite{ref11}, and of the formation
of the intergalactic medium. Related to this, there is a prediction of a new
kind of astrophysical entity: massive bodies which can appear to emit
gamma rays (and X-rays) approximately continously, at low luminosities. We
estimate the total amount of dark matter that may be present in those
discrete massive entities which are related to gamma-ray bursts; it is less
than 1\% of critical density. The dynamics based upon energy in circulating
neutrinos, naturally allows for a moderate degree of collimated ejection of
matter and radiation (beaming) in some gamma-ray bursts. Finally, we consider
some general consequences of the dynamical idea of energy in circulating
neutrinos, for dark-matter entities both less massive, and more massive,
than those associated with gamma-ray bursts. An example of the
latter is in M87, an active galactic nucleus of relatively low luminosity,
which gives evidence of association with a very large central mass, 
$\sim 3\times 10^9 m_\odot$.\cite{ref12} For quasar-like entities \cite{ref7},
the possible total energy from high-energy neutrinos is secondary to the
gravitational energy released by infalling matter. In active galactic 
nuclei and quasars, the neutrino energy allows naturally, for the
outward jetting of some energy. For binary systems suspected of
harboring a compact entity like a black hole of $\sim 10 m_\odot$,
which are observed as X-ray transients \cite{ref12}, the luminosity
in neutrinos might provide a dynamical mechanism for maintaining
a region of very energetic, but diffuse and poorly radiating
material, near to the compact, massive entity, in long time
intervals between the X-ray outbursts.
\section{Sufficient total energy for GRB, and natural time scales
for duration and fluctuation}
Consider a massive body $M$, of about $3.3\times 10^6 m_\odot$ ( $m_\odot$ is
the solar mass, $\sim 1.1\times 10^{57}\GeV$). Assume that the body,
and its immediate vicinity, contains primarily about $3\times 10^{52}$
dark-matter particles, inflatons of mass calculated to be about \cite{ref6}
$10^{11}\GeV$. $^{F1}$ Assume that the state of the body is close
to that of a Schwarzschild black hole \cite{ref10}, that is with a
boundary at a radius $r\simgt r_{\mathrm S}  = 2GM \cong 10^{12}\cm$.
The particular approximate size of this dimension plays an important
role in the dynamics described below.
Inflatons may decay into $\nu_\tau + \ol{\nu}_\tau$, with a lifetime
estimated to be about $10^{25}$ s. \cite{ref6,ref7} As stated in the
introduction, we invoke the assumption that a significant number
(idealized here as most) of these energetic neutrinos are spatially
confined: they circulate and accumulate in a spherical envelope,
in the immediate vicinity of the boundary. We use for this dimension $r\simgt 2\times 10^{12}\cm$,
and use a spread of $2\Delta r\sim 10^{12}\cm$ (see the appendix)
in each
of the two directions perpendicular to the (instantaneous) tangent
to any great circle on the encompassing spherical envelope. The
total energy in all of the decay neutrinos moving in great circles
at some time $t$, (which is initially measured from $t\cong 0$, where
approximately none are present), is proportional to $(t/\tau)$, and is in this example
\be
E_{\mathrm total} \cong (3.3\times 10^{52}\;\mathrm{inflatons}) 
\times (10^{11}\GeV) \times \left(\frac{t_\nu}{10^{25}\s}\right)
\sim 10^{52}\ergs
\ee
for $t_\nu\sim 0.17\times 10^{17}\s \sim 0.038t_{\mathrm{U}}$, where
$t_{\mathrm{U}}$ is the present age of the universe, taken as $\sim 
4.5\times 10^{17}\s$. As explained in detail in section 4, the time
interval $t_\nu$ is approximately determined by the requirement
that the density of the accumulating neutrinos is still below that
at which self-interaction occurs with large probability (and hence,
energy dissipation through created charged particles).
\footnote{In the order-of-magnitude
estimate in Eq.~(1), gradual loss of neutrinos due to self-interaction
up to $\sim t_{\nu}$, is not taken into account.
}  
\footnote{We do not explicitly consider here, effects of ``climbing out''
of a gravitational potential from $r\sim 3r_{\mathrm{S}}$,
since we are dealing with order-of-magnitude estimates, for which we also use
a maximum energy in decay neutrinos.} 
If initially measured from the universe ``origin'',
the interval $t_\nu$ corresponds to a red-shift $z\sim 8$.

Consider that a toroidal configuration of infalling matter rapidly encounters
the circulating neutrinos, at $\sim t_\nu$. A possible density\footnote{The matter is
not extremely dense, as in some ``coellescence'' models\cite{ref2}, where
densities of $10^{(33-36)}\cm^{-3}$ are considered, because dimensions
of $\simlt 10^8\cm$ are invoked. These models often utilize neutrino-antineutrino
pairs to transport energy \cite{ref2} to a much larger dimension
($\simgt 10^{12}\cm$), where the physical processes involving energized
electrons (and positrons) occur, giving rise to the observed photons.} of such
infalling (ordinary) matter at $r\sim 10^{12}\cm$ might be
$\rho\sim 3\times 10^{20}\cm^{-3}$ (approximated
as scaled up from an estimated $\sim 3\times 10^{11}\cm^{-3}$ at
$\sim 10^{15}\cm$, for a
modelled, self-gravitating protostar\cite{ref13}). The last time
interval for free-fall of the matter to the central entity
would probably be only of the order of a few minutes, similar
to the burst duration time, estimated in the paragraph below.
The infall depletes most of the matter.
The dynamical
time interval for interaction within the envelope of circulating
high-energy neutrinos, is given by ( $c$ is the speed of light )
\be
\delta t\sim \frac{\ell_\nu}{c} = \frac{1}{c \sigma_\nu\rho}\cong 1\s
\ee
We have used an extrapolated \cite{ref14} $\sigma_\nu\sim 10^{-31}\cm^2$,
appropriate to such extremely high-energy neutrinos.
The fact that this cross section is naturally large, relative to that
for MeV neutrinos, plays a significant role in the dynamics here.
The time interval in Eq.~(2) provides a natural basis for considering
the empirical, highly irregular fluctuations in intensity, within
given GRB. For
example, a density variation within the toroidal matter, say up 
by a factor of $\sim 10$, can give a time interval for a marked 
upward fluctuation of $\sim 0.1\s$. \footnote{Assuming that the fluctuation in density
is ``cleared'' in about the corresponding neutrino-interaction time.}. 
Little density variation
results in little fluctuation over the burst duration time.

We estimate the full, burst duration time. At $\rho\sim 3\times 10^{20}\cm^{-3}$,
the approximate toroidal volume of $\sim 2\pi r(2\Delta r)^2\sim
1.26\times 10^{37}\cm^3$, contains $\sim 3.8\times 10^{57}$ particles
(the equivalent of $\sim 4m_\odot$). Within $\Delta t\sim 2\Delta r/c \sim
33\s$, the matter encounters $\sim 2\pi r(2\Delta r)^2/(4\pi r^2(2\Delta r))=
\Delta r/r$ of the neutrinos. The time for encountering all neutrinos
in the spherical envelope is the approximate duration time $t_{\mathrm{dur}}
\sim (2\Delta r/c)(r/\Delta r)
= 2r/c\sim 130\s$. Thus, a sensible duration time arises naturally. We have
illustrated this with a definite preferred numerical example (see the
appendix). If one considers a smaller massive body, say with $M$
reduced by $\sim 10^{-2}$, the duration time will be of the order of 1 s.
However, the total energy will be reduced, to less than $10^{50}\ergs$, because
of the smaller number of accumulated high-energy neutrinos from decay
of the massive dark matter.

It is useful to briefly compare the above dynamics for $\delta t$ and $t_{\mathrm{dur}}$
to that of common models \cite{ref15}. In these models, $\delta t$ arises
as $R_{\mathrm{source}}/c$, where $R_{\mathrm{source}}$ is the radius of some
unknown ``internal engine'', usually set at $\sim 10^8\cm$. However,
the dimension at which the observed gamma rays are emitted is necessarily \cite{ref15}
much larger, $\sim \gamma^2 R_{\mathrm{source}}\sim 10^{12}\cm$ to $\sim 10^{14}\cm$,
where $\gamma\sim 10^2$ to $\sim 10^3$ is an initial, extremely large Lorentz factor
for the bulk (i.~e.~coherent) motion of a rather limited, definite number
of baryons $(\sim 10^{52})$ in a ``shell'' of dimension $\Delta$. The duration
time is $\Delta/c$, the time that the ``inner engine'' is active. The
dynamics by which such a large bulk $\gamma$ is brought about for the shell,
is unknown. The reason for a limited number of energized baryons is
unknown. The kinetic energy of the shell must be ``thermalized'' in
energetic electrons, by internal mechanisms \cite{ref15}, before emission
of the gamma-rays. This process is not particularly efficient \cite{ref16,ref4},
and leads to the prediction that there must be a large fraction of the
total energy in the kinetic energy of the ejecta, and hence in ``afterglow'' radiation.\cite{ref16}
As we indicate in the
next section, the present ideas lead to the opposite prediction. The
necessary energy in electrons is here efficiently provided by the interaction
of very energetic neutrinos with infalling matter. These energetic 
electrons emit photons in rapid interaction with the matter, 
and the electrons Compton up-scatter lower energy photons (as is discussed
in detail in sections 3 and 5).
\section{Smaller total energy in ejecta, and in prompt X-rays from infalling matter}
The $E_{\mathrm{total}}$ in Eq.~(1) involves $\sim 10^{44}$ neutrinos
($\nu$ and $\ol{\nu}$). Collisions in a torodial configuration 
of infalling matter, energize electrons, and nucleons. Much of the primary available
energy in the neutrinos goes into electrons; these undergo many subsequent
electromagnetic interactions with the matter, over a path length of the order of $10^{12}\cm$. 
A mean-free path of $\sim (1/\sigma_{\mathrm em}\rho)\sim 10^5\cm$,
allows for $\sim 10^{12}/10^5 = 10^7$ such interactions (using an approximate
cross section $\sigma_{\mathrm{em}}
\sim 10^{-25}\cm^2$, as discussed in section 5). Thus, there are effectively
$\sim 10^{44}\times 10^7=10^{51}$ highly energized individual electrons. 
Assuming a comparable
number of energized nucleons, with individual Lorentz factors of up to
$\gamma_p\sim 10^3$, gives a total baryonic kinetic energy of up to about $10^{51}\ergs$,
which is about $10\%$ of the energy in gamma-rays (the approximate, neutrino
total energy in the example in Eq.~(1)).
A bulk motion of all these nucleons cannot give more (even for hypothetical bulk
$\gamma$-factors of up to $\sim 10^3$). 
Clearly, the present dynamics obviates the situation in which a very energetic ``shell'',
composed of an (apriori) limited number
of baryons, must transfer energy to electrons. Here, extremely energetic electrons
interact with infalling matter, transferring some amount of kinetic energy to a 
limited number ($\sim 10^{51}$) of baryons.
The limited number arises naturally from the limited number of decay neutrinos. 
The prediction is clearly that the total
kinetic energy of the baryonic ejecta is only a small fraction of the total
energy in gamma rays.

The infalling matter which encounters the circulating neutrinos involves of the
order of $4\times 10^{57}$ particles. The release of gravitational energy
as electromagnetic radiation from these infalling particles might be roughly estimated
as thermal X-rays\cite{ref17}, with a luminosity given by
\be
{\cal{L}} = 4\pi r_{\mathrm S}^2\sigma T^4\sim 10^{49}\ergs \s^{-1}
\ee
for $r_{\mathrm{S}}\sim 10^{12}\cm$ and temperature $T\sim 10^7 \;^\circ{\mathrm{K}}$, corresponding
to prompt X-rays of about 1 keV. ( $\sigma$ is the Stefan-Boltzmann constant.)
A similar estimate for ${\cal{L}}$,
follows from a rate of mass infall of about $(10^{-2}m_\odot)\s^{-1}$,
with $\sim 0.1$ \% efficiency for conversion of mass to electromagnetic
radiation.
Over $\sim 100\s$, this total energy
is about $10^{51}\ergs$; so less than the total
energy in gamma rays. Some of these photons can be subsequently Compton
up-scattered by the very energetic electrons produced by the interactions
of the circulating neutrinos.\footnote{
The number of these photons can be initially comparable to the number 
of gamma-rays. The burst emission of X-rays and gamma rays overlap in time.
X-rays can preceed and can follow the main bursts of gamma rays. A decline
of prompt X-rays relative to gamma rays is expected.} 
Of the order of $10^7$ such
interactions initiated by each of $\sim 10^{51}$ highly-energized electrons,
would produce of the order of $10^{58}$ gamma rays, with the approximate
burst $E_{\mathrm{total}}$ of $10^{52}\ergs$ (as in Eq.~(1)).
\section{Significant spread in total energies; presence of GRB at very
large red-shifts; and existence of entities with approximately continuous
emission of gamma rays}
The accumulation of high-energy neutrinos in the immediate vicinity of a 
massive, dark-matter entity must be limited by the self-interaction of
the neutrinos. From a time set as $t\cong 0$, their density from
inflaton decay grows as approximately $(t/\tau)$. Therefore, a rough
estimate of a possible accumulation time interval follows from
\be
t_\nu\sim \frac{1}{c\rho_\nu(t_\nu/\tau)\rho_{\mathrm{inf}}} \rightarrow
t_\nu \sim \sqrt{\frac{\tau}{c\sigma_\nu\rho_{\mathrm{inf}}}} \cong
0.17\times 10^{17}\s \cong 0.038 t_{\mathrm{U}}
\ee
where we have used an approximate inflaton number density of $\rho_{\mathrm{inf}}
\sim 10^{14}\cm^{-3}$, and $\sigma_\nu\sim 10^{-32}\cm^2$, with $\tau\sim
10^{25}\s$ an approximate lifetime for decay into $\nu_\tau\ol{\nu}_\tau$ of massive
dark matter. The time interval estimated in Eq.~(4) has been used in Eq.~(1)
to estimate $E_{\mathrm{tot}}$. $^{F4}$ Accumulation of neutrinos is
disrupted after $\sim t_\nu$, and the accumulated energy is ``dissipated''
into other particles. For a GRB to occur, around $t_\nu$, infalling
particles in a toroidal configuration of matter must ``suddenly'' encounter the
circulating neutrinos, very near to the central entity. An initial interval
of $\sim t_\nu$, corresponds to a red-shift of $\sim 8$. This suggests that
GRB have been prolific at very early times in the universe. The distribution
of $t$ about $\langle t\rangle \sim t_\nu$ is very wide: the dispersion is 
$\sim 0.5\langle t\rangle$. This means that the intrinsic distribution of total
energies from gamma-ray bursts is wide; in particular spreading to values
well below $\sim 10^{52}\ergs$. The present dynamics clearly favors the occurrence
of gamma-ray bursts at early times, in local regions of dense matter, that is
in regions which may exhibit elevated star-formation activity.
Such a locality is not necessarily in a galactic center. At later times,
as the universe thins out, GRB become much less common, in particular in
the present epoch.

After a time interval $\simgt t_\nu$, the process of neutrino accumulation,
and subsequent dissipation through self-interaction, can continue to occur
near to the massive, dark-matter entity. If no abrupt encounter with sufficiently
dense matter
occurs, the system could appear \footnote{Gamma rays result from hadronic
fragments in neutrino-antineutrino interactions, and from interactions of
the neutrino-produced, very high energy electrons, with matter along
the line-of-sight to the source.} as an approximately continuous emitter
of gamma-rays (and X-rays), with a luminosity building up to of the order of\footnote{
If the emission were to occur over only a relatively short time interval,
say of the order of 100 years, a luminosity of $\sim 10^{10}$ times the solar
luminosity could appear from the immediate vicinity of a compact, massive
entity.}
\be
{\cal{L}} \sim 10^{52}\ergs/0.17\times 10^{17}\s \sim 6\times 10^{35}\ergs\s^{-1}
\ee
Such entities could exist in our local environment in the present universe,
where the conditions for GRB to occur have become much less probable; observable
in particular, in regions of space where gas clouds occur between the entity
and the Earth, allowing very high-energy electrons from the source to
interact and radiate, around the line-of-sight.

Two questions arise: what is the possible contribution of dark matter in such
entities to the matter density of the universe, and what flux of extremely
high-energy neutrinos might arise at the Earth (in cosmic-ray, air shower
experiments), from a discrete entity which is not too distant? To roughly
estimate the first from the number of gamma-ray bursts,
we use a GRB rate of $\sim 300$ per year over $\sim 10^{10}$ years,
so $\sim 3\times 10^{12}$ GRB. With a mass of the dark-matter entity of
$\sim 3\times 10^6 m_\odot$, the contribution to the mass density
out to $\sim 10^{28}\cm$ (in an equivalent ``diffuse'' distribution) is
\be
\rho_M \sim \frac{10^{19}\times 10^{57}\GeV}{\frac{4\pi}{3}\times 10^{84}\cm^3}
\sim 0.25\times 10^{-8}\GeV\cm^{-3}
\ee
This is only $\sim 0.05\%$ of a critical density of about $0.5\times 10^{-5}\GeV\cm^{-3}$.
It is therefore well below limits for the contribution from such massive entities,
which are claimed by recent experiments involving gravitational 
lensing \cite{ref18,ref19}. Using the luminosity in Eq.~(5), and assuming that of
order of one-quarter of the energy is radiated in the highest-energy neutrinos
of $\sim 10^{20}\eV$, results in an estimated flux from a discrete entity at
a distance of $\sim 4\times 10^{22}\cm$, of\footnote{If there is some beaming
(as in section 6), this flux is for beaming along the line-of-sight.}
\be
f_\nu\sim\frac{2\times 10^{38}\GeV\s^{-1}}{10^{11}\GeV (4\pi\times(4\times 10^{22}\cm)^2)}
\sim 10^{-19}\cm^{-2}\s^{-1}\sr^{-1}
\ee
Upcoming\cite{ref20} air-shower experiments with cosmic rays will search for a flux
of neutrinos at $\sim 10^{20}\eV$, in particular through the presence of showers
initiated at large zenith angles.\footnote{Approximated as a ``diffuse'' distribution
of discrete sources out to $\sim 10^{28}\cm$, $\sim 10^{12}$ entities can give a flux
of the order of $10^{-18}\cm^{-2}\s^{-1}\sr^{-1}$ for neutrinos which are only
moderately red-shifted to lower momenta (again, assuming little reduction
of observable flux due to the effect of beaming).}
\section{Photon emission processes}
It is clear that the hypothetical dynamics discussed in this paper, allows
for a prolific source of extremely high-energy electrons (and positrons).
Thus, Compton up-scattering of photons must be an important source of
gamma rays. There are many discussions \cite{ref15} which
carry a strong dependence upon synchrotron radiation, under the assumption of
the existence of sufficiently strong magnetic fields, in particular at early
times. There exists some data on photon polarization \cite{ref21}, which
raises the question as to its consistency with the usual assumption concerning
large magnetic fields, with spatial extension. It is well 
known \cite{ref22} that quite large magnetic
fields are necessary in order to get sufficiently short characteristic times
$t_{\mathrm{sync}}$, for synchrotron radiation. For example, for an electron
of 1 GeV, a magnetic field $B\sim 3\times 10^4$ Gauss is required in order
to have $t_{\mathrm{sync}}\sim 3\times 10^{-4}\s$ (and an energy 
$\hbar\omega_{\mathrm{sync}}\sim 2\keV$ ). For $B\sim 3$ Gauss,
$t_{\mathrm{sync}}$ becomes $\sim 3\times 10^4\s$. 

In the present dynamical situation, it also seems likely that bremsstrahlung
should be an important radiation process, providing photons which subsequently
may be Compton up-scattered. For an electromagnetic cross section $\sigma_{\mathrm{em}}$
of the order of $10^{-25}\cm^2$, a dynamical time of the order of $(1/c\sigma_{\mathrm{em}}\rho)
\sim 3\times 10^{-6}\s$ occurs in an inital matter density $\rho\sim 10^{20}\cm^{-3}$.
This is a shorter time interval than $t_{\mathrm{sync}}$, even assuming the presence
of the above large magnetic field. Out to $\sim 10^{15}\cm$, where the matter
density is probably decreased by about nine orders of magnitude, the dynamical
interaction time of $\sim 3\times 10^3\s$ is still shorter than the transit time of
$10^{15}\cm/c = 3\times 10^4\s$. Therefore, independently of the role
of synchrotron radiation in some time intervals of the burst evolution,
and in the emission of radiation at later times (in particular, radio waves), 
bremsstrahlung and Compton up-scattering should be important processes
at early times.
\section{Natural possibility for dynamical beaming of matter and energy}
The dynamics involves the interaction of a toroidal configuration of infalling
matter, with high-energy neutrinos that are traversing great circles
on a spherical envelope which encompasses the immediate vicinity of a compact,
massive entity. This situation provides dynamical and geometrical conditions
in which a moderate degree of beaming of matter and radiation could take place
in some gamma-ray bursts. If the axis of the torus is close to the line-of-sight,
a maximal observed flux is possible; otherwise the observed flux is reduced.
This is a ``geometrical'' effect which, in itself, gives rise to a broad distribution
in observed total energies of gamma-ray bursts, in addition to the intrinsic
broadness which we discussed in section 4. The possibility for beaming goes
away as the ``covering'' angle of the semi-torus on the sphere increases (toward
$\pi$, i.~e.~no toroidal configuration). On the other hand, if the efficiency for reaching
maximum luminosity is not to be reduced, the angle should not be less than of
the order of $\Delta r/r \sim 1/4$. This suggests
the possibility of a moderate degree of beaming characterized by an angle $\theta\sim 1/4$; 
thus a possible reduction to an ``intrinsic'' total energy by
a factor of $(\theta^2/2)\sim 1/32$, from that deduced when isotropic
emission is assumed. In contrast to the ``standard'' model \cite{ref15,ref16},
we have not invoked the standard assumption that ``relativistic beaming''
of bulk matter involves effective, initial $\gamma$-factors of 100 to 1000. Rather, the
possibility is present that such relativistic factors for approximate bulk motion 
may acquire only moderate, more probable values \cite{ref4}, 
say up to $\sim 5$, in the initial burst time
interval. Then, $\theta>\gamma^{-1}$ is not necessarily always satisfied. This is the
usual condition for the possibility of marked observable effects \cite{ref23},
as the bulk matter slows down at  later times, and $(\gamma(t))^{-1}$ changes
from less than $\theta$ to greater than $\theta$. If an effective $\gamma^{-1}$
is greater than $\theta$, then the beaming occurs within the observable
angular interval around the line-of-sight (i.~e.~the relativistic beaming
from approximate bulk motion, into $\sim \gamma^{-1}$), 
already in the initial time interval.
\section{Aspects of the hypothesis of less massive, and more massive, dark-matter
entities}
There are observations \cite{ref12} that concern less massive entities, of the
order of $10m_\odot$, which are hypothetically near to the condition of a black
hole. In particular, observations of binary systems involving such an
entity and a low-mass companion star, referred to as low-mass, X-ray transients
(LMXT). \cite{ref12} These systems sporadically burst into X-rays, and sometimes
repeat, after relatively long time intervals of quiescence i.~e.~periods of approximately
steady, but low luminosity. Models \cite{ref12} for LMXT often involve a (repeatable)
instability in an ``outer, cold'' accretion disk, which results in an  elevated
rate of inflow of matter for a relatively short time, toward the compact,
massive entity. An essential dynamical aspect \cite{ref12,ref24}, is the
hypothetical existence, during the long periods of quiescence, of a spherical,
inner configuration between the accretion disk and the central entity. This region
is supposed to consist of a highly energetic, but dilute, plasma (it can contain
a ``hot corona'' near the boundary to the outer disk). This inner configuration, 
which envelopes the central entity, is a poor radiator \cite{ref12,ref24}. The result
is a low luminosity of radiation emitted by any rapidly infalling matter, during times other
than the burst intervals. An open question \cite{ref12} concerns what energy source gives
rise to this spherical, inner configuration. In particular, after a burst,
what is the dynamical mechanism through which the physical condition of the inner
region suitable for a subsequent burst, is set up? A certain possible similarity
to the GRB model that we have discussed in this paper, lies in the ``sudden''
elevated infall of dense matter through a ``dilute'' spherical envelope of
energetic particles, which encompasses the compact, massive entity. 
However, the energy for the X-ray outburst comes largely from the release
of gravitational energy by the elevated amount of infalling matter, over the burst
time. Our main point here, is that again assuming the presence of high-energy neutrinos
in the immediate vicinity of this entity, then a minimum, nearly steady luminosity
can occur during long quiescent time intervals; we estimate this. The number of
inflatons of calculated \cite{ref6} mass $\sim 10^{11}\GeV$, to approximately
account for an entity of $\sim 10 m_\odot$ is $\sim 10^{47}$. A comparable
number probably exists in the immediate vicinity of the surface, perhaps with
significant rotational motion. Inflatons may decay into neutrino-antineutrino
pairs, with an estimated lifetime \cite{ref6,ref7} of $\sim 10^{25}\s$. The
high-energy neutrinos can interact with matter at the inner edge of
the outer disk, energizing electrons and producing an approximately steady
luminosity of the order of
\be
{\cal{L}}_{\mathrm{quiescent}} \sim 10^{47}\times 10^{11}\GeV/10^{25}\s \cong
1.6\times 10^{30}\ergs \s^{-1}
\ee
This minimum luminosity may often be exceeded by that radiated (with anomalously
low efficiency \cite{ref24,ref12}) by the matter, during the short time
of infall. In any case, the level of luminosity during quiescence is far below
that which occurs during the X-ray outburst ($\sim 10^{38}\ergs \s^{-1}$). \cite{ref12} 
We estimate that the minimum, steady luminosity in Eq.~(8) is sufficient to ``clear''
an (elevated) density of ordinary matter of $\sim 10^{19}\cm^{-3}$, from
a region in the immediate vicinity of the massive entity, over a time interval
of months. This time interval is comparable to observed effective burst times
for LMXT (presumably related to the duration of the instability-induced, elevated
mass-flow from the outer disk). \footnote{In an intermediate
mass range between LMXT and GRB, say $3\times 10^3 m_\odot$, there might be a large
number of compact, dark-matter entities. The dark matter in the universe
might then be more discretely compounded, than is usually assumed to be the case.
Gravitational lensing might eventually be able to reveal entities in this mass
range.}

It is worth noting that a well-known, active galactic nucleus (AGN) in the nearby,
radio galaxy M87, may contain a very large central mass, estimated as 
$\sim 3\times 10^9 m_\odot$.\cite{ref12} This system has a luminosity of only
$\sim 10^{43}\ergs \s^{-1}$, and exhibits some collimated ejection perpendicular
to a disk structure, and at a sizable angle to the line-of-sight. The last time
interval for infall of matter ($\sim 3\times 10^4\s$ from $\sim 10^{15}\cm$)
is comparable to a radiation time ($\sim (1/c\sigma_{\mathrm em}\rho)\sim 3\times 10^{4}\s$
for $\rho\sim 10^{10}\cm^{-3}$). This system is thus a very massive candidate
for forming an ADAF-type \cite{ref12} of ``hot, dilute'' inner configuration which envelops
the central mass. Within the framework of the present ideas, this region
could contain a steady luminosity of very high-energy neutrinos, which can energize
the infalling electrons. This naturally allows for the possibility of collimated
ejection of some energy in approximately ``bulk'' matter; the electrons radiate
from the jet. Such entities may well be nearby, as is M87.

Very energetic quasars \cite{ref25}, such as 3C273 at $z\sim 0.16$, are present at
distances from the Earth which are not the greatest. Some quasars are not clearly
in the center of a galaxy. \cite{ref25} Allowing for masses of the 
dark-matter entity of up to \cite{ref7} $\sim 10^{11} m_\odot$, results in a total
energy release from the luminosity of high-energy neutrinos from decay of
dark matter, of less than $10^{58}\ergs$, over about $10^6$ years. Such an energy source would
be secondary to a possible total of $10^{60}\ergs$ (for 3C273), presumably
originating in gravitational energy which is released as radiation from infalling matter
in the vicinity of the massive entity. However, the secondary energy has the natural
possibility of jetting (which is marked in 3C273, where the ratio of radio wave
to X-ray luminosity is about $10^{-2}$). The time for onset of quasar activity, and
its presumed limited duration, is generally related to the occurrence of a sufficient
rate of infalling matter \cite{ref25}; and thus to galactic collisions, and/or
to epochs of elevated star formation. However, in the present model, the
very massive central entity formed of massive, dark-matter particles \cite{ref7}
may well have been present at an earlier time in the universe, after inflation. \cite{ref6}
\section{Summary}
The usual dynamics in the space near to compact massive bodies which are close
to the condition of a black hole, involves the accretion of matter from nearby
bodies, or generally, from a sufficiently dense surrounding medium. In this paper,
we have considered the presence, and the possible accumulation in rotation,
of very energetic neutrinos (and antineutrinos), which arise from the very
slow decay of massive, dark-matter particles that are assumed to be present
in a massive, central body, and in its immediate vicinity. In an appendix,
we argue heuristically for the assumed possibility, but we are uncertain
as to whether it can be realized. \footnote{It is interesting that a recent
paper \cite{ref26} has speculated about an ``extra'' energy source for the solar
corona. The origin of the energy is hypothesized to be neutral, scalar particles
as light as 10 eV. These can be produced \cite{ref27} deep inside the sun itself,
and some are assumed to be gravitationally retained  and accumulated in orbits
in the Sun's vicinity, with no dissipation. These particles are supposed to
decay into two photons (at X-ray energies), with a very long lifetime. However,
it is difficult to gravitationally  constrain light particles, since the escape
velocity from the Sun is relatively low. Note that a flux of extremely high-energy
neutrinos (as in Eq.~(7)) can give rise to an external-energy ``irradiation'',
acting upon the whole Sun, of the order of $10^{13}\ergs \s^{-1}$, more than that
from any hypothetically localized, decaying inflaton matter.}
It is our purpose here, to raise the possibility.
We have shown that a large number of interesting consequences follow. These are
relevant for on-going observations, in particular for gamma-ray bursts. Thus,
further investigation of the possibility would seem to be useful.\footnote{
The new dynamical idea discussed here is speculative. In the absence of
reasons for extraordinary behavior, so are common ideas which postulate the bulk
motion of $\sim 10^{-5} m_\odot$ with $\gamma$-factors of up \cite{ref15} to
1000, and/or the coherent rotation of a massive body \cite{ref4} at nearly
the speed of light, in the presence of enormous magnetic fields. An essential
element in the present ideas can be directly tested in upcoming experiments,
which will be capable of detecting a significant flux of neutrinos in the
highest-energy cosmic rays.\cite{ref20,ref7}}
\section*{Appendix} 
\def\theequation{A\arabic{equation}}
\setcounter{equation}{0}
Neutrinos have mass \cite{ref8,ref9}. The largest mass, probably that of $\nu_\tau$,
may be about 0.06 eV. The smallest mass, presumably that of $\nu_e$, may
not be less than a factor of 0.1 smaller. Large flavor mixing can occur,
over sufficiently long flight paths (which lengthen with increasing
neutrino energy). There may occur (meta)stable orbits \cite{ref10}
for neutrinos with large angular momenta, out from about $3/2\times 10^{12}\cm$ around
a compact, massive entity, $M\sim 3\times 10^6 m_\odot$, which is near to
a black-hole condition ($r_{\mathrm{S}}\sim 10^{12}\cm$). It might be possible
for these to accumulate over long periods of time, in an envelope over
the sphere. A heuristic argument is as follows. Consider as measure of the
``degree of stability'' at some $r$, the quantity\footnote{
The factor $(1/m_\nu r)$ is a maximal dimensionless parameter, which arises
from differentiations of the effective potential given in the caption
to Fig.~19 on p.~67 of Ref.~10.}
\be
\Delta r \sim \frac{1}{m_\nu r} \times (c t_{\mathrm{U}})
\ee
Consider for stability that $\Delta r$ in either of the two directions perpendicular
to an instantaneous tangent to any great circle is limited to be $\simlt r/2$.
Then, the order of magnitude of (a minimum) $r$ is determined by
\be
r \sim \left\{ \frac{2ct_{\mathrm{U}}}{m_\nu}\right\}^{1/2} \sim
\left\{ \frac{2.7\times 10^{28}\cm}{(3.3\times 10^{-4}\cm)^{-1}}\right\}^{1/2}\sim
3\times 10^{12}\cm
\ee
for $m_\nu \sim 0.06 \eV$, $t_{\mathrm{U}}\sim 4.5\times 10^{17}\s$. As estimated
in Eq.~(4), $t_{\mathrm{U}}$ in Eq.~(A2) should be replaced by $t_\nu\sim 0.038 t_{\mathrm{U}}$.
Then, taking a smallest $m_\nu$ as $\sim 0.006\eV$, results in 
$r\sim 2\times 10^{12}\cm$.
This is close to the dimensions of (meta)stable orbits about an entity of mass
$M\sim 3\times 10^6 m_\odot$,
which is near to the condition of a black hole, i.~e.~with a boundary at
$r\simgt 10^{12}\cm$, as used in section 2.

Clearly, for more massive entities in active galactic nuclei and quasars, with
$r$ of the order of $10^{15}\cm$ to $3\times 10^{16}\cm$, the $\Delta r$ from
Eq.~(A1) is $\ll 1$. The stability argument fails for the small values of $r$
relevant to the discussion of LMXT in section 7. However, in this situation
accumulation of neutrinos is not required. The inner ADAF is ``confined''
by the outer disk; that is, the luminosity from the high-energy neutrinos
tends to persistently ``evaporate'' matter at the inner edge of the ``cold''
disk, resulting in a corona of very energetic electrons.

The result in Eq.~(A2) may imply a particular cosmological connection for
neutrino mass. It is noteworthy that the inverse of neutrino masses of the order of
0.06 eV to 0.006 eV, give times of the order of $10^{-14}\s$ to $10^{-13}\s$.
These times are just prior to the approximate time for the breaking of
electroweak symmetry in the early universe, at
$10^{-12}\s$ (energy scale $\sim 1\TeV$). Thus, neutrino mass is, in a sense,
a minimal mass ``uncertainty'' related to this early time interval.
(This suggests that neutrino mass might originate in vacuum energy \cite{ref28}
other than the vacuum-expectation value of a Higgs field.)  

\end{document}